\documentclass[amsmath,pra,twocolumn,showpacs]{revtex4}
\usepackage{graphics}

\begin{document}     

\title{Cooperative enhancement of channeling of emission from atoms into a nanofiber}

\author{Fam Le Kien and K. Hakuta} 
\affiliation{
Department of Applied Physics and Chemistry, 
University of Electro-Communications, Chofu, Tokyo 182-8585, Japan
}

\date{\today}

\begin{abstract}
We show the possibility of directional \textit{guided} superradiance  
from a string of atoms separated by one or several wavelengths in a line parallel to the axis of a nanofiber. We find that
the rate and efficiency of channeling of emission from the atoms into the fiber are cooperatively enhanced by the guided modes. 
\end{abstract}

\pacs{42.50.Fx,42.50.Ct,42.81.-i}
\maketitle

Coupling of light to subwavelength structures and its control pose one of the greatest challenges of recent research \cite{Maier,Wallraff,Lukin,cesium decay,Kali,Nayak}. Strong coupling in a superconducting circuit at microwave frequencies has been observed \cite{Wallraff}. Chang \textit{et al.} have proposed a technique that enables strong coherent coupling between individual emitters and guided plasmon excitations in conducting nanostructures \cite{Lukin}. In the case of \textit{dielectric} waveguides, it has been shown that a significant fraction of emission from a single atom can be channeled into a nanofiber \cite{cesium decay,Kali,Nayak}.
The formation of single-atom trapping sites on a nanofiber surface
without any external field and the coupling of single photons to a nanofiber 
have been demonstrated \cite{Nayak}. 
The cooperation of two distant atoms via a nanofiber has been discussed \cite{two atoms}. It has been shown that, at large distances between the atoms, a substantial energy exchange can survive due to the guided modes \cite{two atoms}. In this paper, we show the possibility of a directional \textit{guided} superradiant emission process that can enhance the rate and efficiency of channeling of emission from a string of distant atoms into a nanofiber.

Before we proceed, we note that superradiance is a problem of fundamental interest \cite{Dicke}. Despite a great deal of research \cite{Gross,Mandel book}, certain aspects of the problem are still not well understood. An example is the mode selection in the directional emission from an extended sample of atoms. The main difficulty is due to the fact that the collective process involves a huge number of degrees of freedom, associated with many atoms and a continuum of field modes. Recently, the angular distribution of emission from a spatially extended array of atoms in free space has been treated by the quantum trajectory approach \cite{Carmichael}. The dynamic mode selection has been studied \cite{Law}. Superradiant conversion of atomic spin gratings into single photons in an optical cavity has been demonstrated \cite{Black}.  

\begin{figure}[tbh]
\begin{center}
  \includegraphics{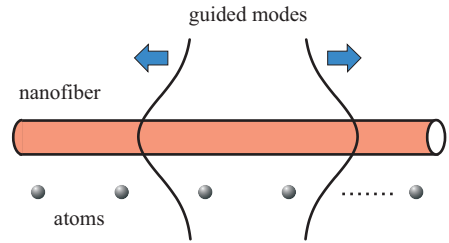}
 \end{center}
\caption{String of atoms in the vicinity of a nanofiber.}
\label{fig1}
\end{figure}

Consider $N$ identical two-level atoms interacting with the quantum electromagnetic field
in the vicinity of a nanofiber (see Fig. \ref{fig1}). 
The fiber has a cylindrical silica core of radius $a$ and refractive index $n_1=1.45$ and an infinite vacuum clad of refractive index $n_2=1$. We assume that the atomic frequency $\omega_0$ is well below the cutoff frequency of the fiber, so the single-mode condition is satisfied for 
this frequency. In view of the very low losses of silica in the wavelength range of interest, we neglect material absorption. The atoms are located at points 
$(r_j,\varphi_j,z_j)$, where $j=1,2,\dots, N$ labels the atoms
and $(r,\varphi,z)$ is the cylindrical coordinates with $z$ being the axis of the fiber.
We assume that the field is initially in the vacuum state. 
The field can be decomposed into the contributions from the guided and radiation modes, 
whose quantum expressions are given in  Ref.~\cite{cesium decay}.

Assume that the characteristic atomic lifetime is 
large as compared to the optical period $2\pi/\omega_0$ and to 
the light propagation time between two different atoms.
The master equation for the reduced density operator $\rho$ of the atomic system
in the electric-dipole, rotating-wave, and Born-Markov approximations
has been previously derived \cite{Gross,two atoms,Carmichael,Agarwal}. In the interaction picture, it reads 
\begin{equation}\label{13a}
\dot{\rho}=\frac{1}{2}\sum_{i,j=1}^{N}\gamma_{ij}(
2\sigma_{j}\rho\sigma_{i}^{\dagger}
-\sigma_{i}^{\dagger}\sigma_{j}\rho-\rho\sigma_{i}^{\dagger}\sigma_{j}).
\end{equation}
Here, $\sigma_{j}$ and $\sigma_{j}^\dagger$ are the pseudospin operators that 
describe the downward and upward transitions of the atoms. 
The coefficients
$\gamma_{ij}=\gamma_{ij}^{(\mathrm{guided})}+\gamma_{ij}^{(\mathrm{rad})}$,
with $i,j=1,2,\dots,N$, 
characterize the collective spontaneous emission process, where 
$\gamma_{ij}^{(\mathrm{guided})}$ and $\gamma_{ij}^{(\mathrm{rad})}$
are the contributions from the guided and radiation modes, respectively \cite{two atoms}. 
In Eq. (\ref{13a}), we have neglected the propagation effects and the dipole-dipole interactions. For these  approximations to be valid, the sample must not be too large  
and the distances between the atoms must not be too small. 
We note that Eq. (\ref{13a}) has the same form as that for atoms in free space \cite{Carmichael,Agarwal}.
  
We introduce the total emission intensity $I\equiv\sum\hbar\omega_{\alpha}\langle\dot{n}_{\alpha}\rangle$, the intensity of emission into the guided modes 
$I_{\mathrm{guided}}\equiv\sum_{\mathrm{guided}}\hbar\omega_{\alpha}\langle\dot{n}_{\alpha}\rangle$, and the intensity of emission into the radiation modes 
$I_{\mathrm{rad}}\equiv\sum_{\mathrm{rad}}\hbar\omega_{\alpha}\langle\dot{n}_{\alpha}\rangle$. 
Here, $\omega_{\alpha}$ and $\langle n_{\alpha}\rangle$ are the frequency and
mean number, respectively, of photons in mode $\alpha$. We find 
$I=\hbar\omega_0\sum_{ij}\gamma_{ij}
\langle \sigma_i^\dagger\sigma_j\rangle
=I_{\mathrm{guided}}+I_{\mathrm{rad}}$,
where
$I_{\mathrm{guided}}=\hbar\omega_0\sum_{ij}\gamma_{ij}^{(\mathrm{guided})}
\langle \sigma_i^\dagger\sigma_j\rangle$
and 
$I_{\mathrm{rad}}=\hbar\omega_0\sum_{ij}\gamma_{ij}^{(\mathrm{rad})}
\langle \sigma_i^\dagger\sigma_j\rangle$.
We note that $I=-\hbar\omega_0 \dot{P}$, where
$P=\sum_j\langle\sigma_j^\dagger\sigma_j\rangle$ is
the total population of the excited levels of the atoms.
The total energy emitted from the atoms is 
$U= \int_0^{\infty} I(t)\,dt=U_{\mathrm{guided}}+U_{\mathrm{rad}}$,
where $U_{\mathrm{guided}}= \int_0^{\infty} I_{\mathrm{guided}}(t)\,dt$ and 
$U_{\mathrm{rad}}= \int_0^{\infty} I_{\mathrm{rad}}(t)\,dt$ are the energies emitted into the guided and radiation modes, respectively. The fractions of energy emitted into the guided and radiation modes are given by
$f_{\mathrm{guided}}=U_{\mathrm{guided}}/U$ and $f_{\mathrm{rad}}=U_{\mathrm{rad}}/U=1-f_{\mathrm{guided}}$, respectively.

The diagonal coefficients $\gamma_{jj}$ describe the spontaneous decay of individual atoms.
The off-diagonal coefficients $\gamma_{jj'}$, with the convention $j\not=j'$, characterize the energy transfer between two atoms. According to Ref. \cite{two atoms},
the contribution $\gamma_{jj'}^{(\mathrm{guided})}$ of the guided modes to the transfer rate 
is periodic in the $z$ direction with the period $\lambda_F=2\pi/\beta_0$, where $\beta_0$
is the longitudinal propagation constant of the guided modes at the atomic frequency 
$\omega_0$. Meanwhile, the contribution $\gamma_{jj'}^{(\mathrm{rad})}$  of the radiation modes reduces to zero with increasing interatomic distance $|z_j-z_{j'}|$. Therefore, in the limit of large $|z_j-z_{j'}|$, the transfer coefficient $\gamma_{jj'}$ is mainly determined by the contribution $\gamma_{jj'}^{(\mathrm{guided})}$ of the guided modes and is almost periodic with the spatial period $\lambda_F$.

\begin{figure}[tbh]
\begin{center}
  \includegraphics{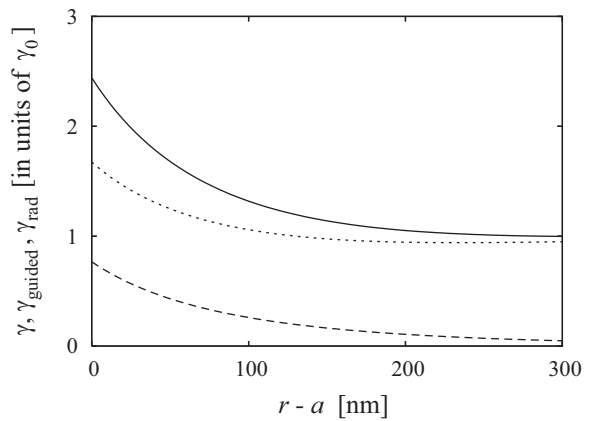}
 \end{center}
\caption{
Total decay rate $\gamma$ (solid line) and the contributions $\gamma_{\mathrm{guided}}$ (dashed line) 
and $\gamma_{\mathrm{rad}}$ (dotted line) from the guided and radiation  modes, respectively, 
as functions of  the atom--surface  distance $r-a$. 
The fiber radius $a=200$ nm and the atomic wavelength $\lambda_0=852$ nm are used.
The dipole of the atom is radially oriented.
}
\label{fig2}
\end{figure}

We now assume that the atoms are aligned along a line parallel to the fiber axis, with relatively large atomic separations being equal to integer multiples of the longitudinal
wavelength $\lambda_F$. 
In other words, we assume that $r_j=\mbox{const}\equiv r_0$, 
$\varphi_j=\mbox{const}\equiv \varphi_0$,
and $z_{j+1}-z_j=q_j\lambda_F$, with $q_j$ being nonzero, positive integer numbers.
In addition, we assume that the dipoles of the atoms are oriented in the same direction.
Under these conditions, the guided transfer coefficients $\gamma_{jj'}^{(\mathrm{guided})}$ achieve their axial maximum value,
$\gamma_{jj'}^{(\mathrm{guided})}=\gamma_{jj}^{(\mathrm{guided})}=\gamma_{j'j'}^{(\mathrm{guided})}$. Meanwhile, due to the large separations between the atoms, the radiative (unguided) transfer coefficients $\gamma_{jj'}^{(\mathrm{rad})}$ are small. In this case, we have $\gamma_{jj}=\gamma_{jj}^{(\mathrm{guided})}+\gamma_{jj}^{(\mathrm{rad})}=\gamma=\gamma_{\mathrm{guided}}+\gamma_{\mathrm{rad}}$ and $\gamma_{jj'}\cong \gamma_{jj'}^{(\mathrm{guided})}=
\gamma_{jj}^{(\mathrm{guided})}=\gamma_{j'j'}^{(\mathrm{guided})}=\gamma_{\mathrm{guided}}$. 
Here, $\gamma_{\mathrm{guided}}$ and $\gamma_{\mathrm{rad}}$ are the rates of decay into the guided and radiation modes,
respectively. They do not depend on the axial coordinate $z$ of the atoms, but increase with decreasing atom--surface distance $r-a$. We display in Fig. \ref{fig2} the rates $\gamma_{\mathrm{guided}}$, $\gamma_{\mathrm{rad}}$, 
and $\gamma=\gamma_{\mathrm{guided}}+\gamma_{\mathrm{rad}}$, calculated for a radially oriented dipole with the cesium $D_2$-line transition wavelength $\lambda_0=852$ nm in the presence of a fiber with radius $a=200$ nm. 
In particular, for the atom--surface distance $r-a=100$ nm, 
we obtain $\gamma_{\mathrm{guided}}=0.26\gamma_0$ and $\gamma_{\mathrm{rad}}=1.06\gamma_0$. Here, $\gamma_0$ is the atomic natural linewidth, whose magnitude is about $5.3$ MHz in the case of the cesium $D_2$ line. We note that the creation of regular strings of atoms in a standing wave optical dipole trap has been demonstrated \cite{Meschede}. 
The formation of single-atom trapping sites on a nanofiber surface
without any external field has been reported \cite{Nayak}. Superradiance of lines of atoms in free space has also been studied \cite{Carmichael}. We emphasize that the transfer coefficients $\gamma_{jj'}$ for distant atoms considered here are mainly due to the guided modes and are substantially larger than those for distant atoms in free space \cite{Carmichael,Agarwal}. 

For the string of atoms described above, we find that the intensity of emission into the radiation modes is 
$I_{\mathrm{rad}}=\hbar\omega_0\gamma_{\mathrm{rad}}P$
and, hence, the intensity of emission into the guided modes is 
\begin{equation}\label{43}
I_{\mathrm{guided}}=-\hbar\omega_0 \left(\frac{dP}{dt}+\gamma_{\mathrm{rad}}P\right).
\end{equation}
Meanwhile, the total excited-state population $P$ is governed by the equation
\begin{equation}\label{23}
\frac{d P}{dt}=-\gamma P-\gamma_{\mathrm{guided}}\sum_{j\not=j'}\langle\sigma_{j}^{\dagger}\sigma_{j'}\rangle.
\end{equation}

\begin{figure}[tbh]
\begin{center}
  \includegraphics{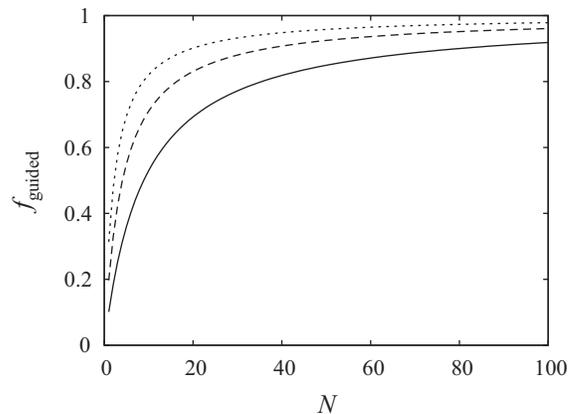}
 \end{center}
\caption{
Fraction $f_{\mathrm{guided}}$ of energy emitted from atoms initially prepared in the symmetric one-excitation state into the guided modes
as a function of the atom number $N$. The atom--surface distance is $r-a=200$ nm (solid line), 100 nm (dashed line),
and 0 (dotted line). 
The parameters used are as in Fig.~\ref{fig2}.
}
\label{fig3}
\end{figure}

We examine two cases of initial atomic states. First we consider the case where
the atomic system is initially prepared in the symmetric one-excitation state
$|1\rangle=N^{-1/2}\sum_j|1_j\rangle$. Here, $|1_j\rangle=|e_j\rangle\otimes 
\prod_{j'\not=j}|g_{j'}\rangle$ is the product state in which only atom $j$ is excited, with
$|e_j\rangle$ and $|g_j\rangle$ being the excited and ground states, respectively, of atom $j$. The state $|1\rangle$ is the first excited Dicke state \cite{Dicke} and is an entangled state, which is of great interest in quantum information and quantum computation 
\cite{entanglement,Duan}. 
We introduce the notation $|0\rangle=\prod_j |g_j\rangle$ for the state in which all the atoms are in their ground states.
We find that the $N$ two-level atoms prepared in the state $|1\rangle$ act like a single effective two-level system,
with the upper level $|1\rangle$ and the lower level $|0\rangle$.
We obtain from Eq. (\ref{13a}) the solution $\rho_{11}=e^{-\Gamma t}$, $\rho_{00}=1-e^{-\Gamma t}$, and $\rho_{10}=\rho_{01}=0$, with the collective decay rate 
\begin{equation}\label{30}
\Gamma=\gamma_{\mathrm{rad}}+N\gamma_{\mathrm{guided}}.
\end{equation}
This rate is enhanced \cite{Dicke} by the cooperativity of the atoms via the guided modes.
The above solution yields the total excited-state population
$P=e^{-\Gamma t}$
and the intensity of emission into the guided modes
\begin{equation}\label{58}
I_{\mathrm{guided}}
=\hbar\omega_0 N\gamma_{\mathrm{guided}} e^{-\Gamma t}.
\end{equation}
Hence the energy emitted into the guided modes is 
$U_{\mathrm{guided}}=\hbar\omega_0 N\gamma_{\mathrm{guided}}/\Gamma$. Meanwhile, the total emitted energy is
$U=\hbar\omega_0$. Consequently,
the fraction of energy emitted into the guided modes is
\begin{equation}\label{60}
f_{\mathrm{guided}}=
\frac{N\gamma_{\mathrm{guided}}}{\gamma_{\mathrm{rad}}+N\gamma_{\mathrm{guided}}}.
\end{equation}
It is clear that $f_{\mathrm{guided}}$ increases with increasing atom number $N$ and that 
$f_{\mathrm{guided}}\to 1$ in the limit $N\to\infty$.
Thus the efficiency of channeling of emission from the atoms into the fiber is cooperatively enhanced. We use Eq. (\ref{60}) to calculate $f_{\mathrm{guided}}$ as a function of $N$ for the parameters of Fig. \ref{fig2}, and display
the results in Fig. \ref{fig3}.
For $N=100$ and $r-a\leq 200$ nm, we obtain $f_{\mathrm{guided}}\geq 0.92$ (see the endpoints of the curves).
In particular, for $N=100$ and $r-a=100$ nm, the factor $f_{\mathrm{guided}}$ reaches the value 0.96 (see the endpoint of the dashed curve). Such a high efficiency indicates that the single photon emitted from the atoms is almost entirely directed into the guided modes. 
A very similar result has been obtained for the superradiance of atoms in an optical cavity \cite{Black}. Indeed, in terms of the single-atom cooperativity parameter $\eta=\gamma_{\mathrm{guided}}/\gamma_{\mathrm{rad}}$, the channeling efficiency $f_{\mathrm{guided}}$ given by Eq. (\ref{60}) coincides with the success probability $\mathcal{P}=N\eta/(1+N\eta)$ for conversion in the cavity case \cite{Black}. Such a coincidence is due to the fact that the nanofiber mode and the cavity mode have many common features. We note that,
at the distance of 100 nm from the surface of the 200-nm-radius fiber,
the single-atom cooperativity parameter is $\eta=0.25$. This value is substantially larger than
the value $\eta=6.9\times10^{-3}$ for a moderate-finesse cavity \cite{Black}.

\begin{figure}[tbh]
\begin{center}
  \includegraphics{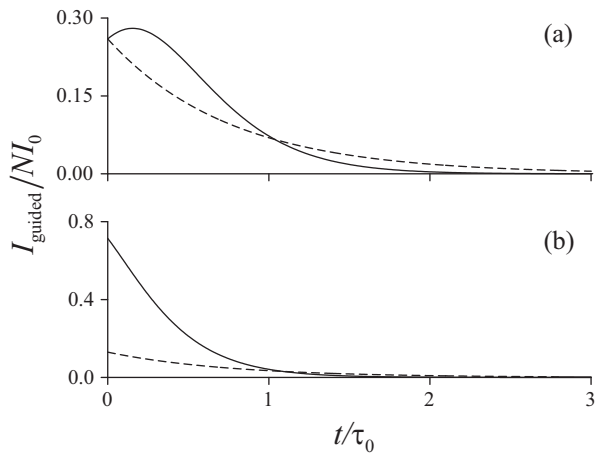}
 \end{center}
\caption{
Intensity of emission from $N=10$ atoms, initially prepared in a product state, into the guided modes as a function of time (solid lines). The angle $\theta$ for
the initial product state is (a) $\theta=0$ (full excitation) and (b) $\theta=\pi/2$ 
(one-half excitation). The distance from the atoms to the fiber surface is $r-a=100$ nm. The intensity is normalized to $NI_0$, where $I_0=\hbar\omega_0\gamma_0$. The time is normalized to $\tau_0=\gamma_0^{-1}$. The parameters used are as in Fig. \ref{fig2}. For comparison, we also show
the results for the case of a single atom (dashed lines).
}
\label{fig4}
\end{figure}

We now consider the case where all the atoms are initially prepared in the same coherent superposition state, that is, the initial state of the atoms is the product state
$|\Psi\rangle=\prod_j (\cos\frac{\theta}{2}|e_j\rangle+e^{i\phi}\sin\frac{\theta}{2}|g_j\rangle)$.
Such a state can be prepared by using a plane-wave optical pulse to excite the atoms.
We solve Eq. (\ref{13a}) numerically for $N=10$ atoms, use this solution to calculate
the intensity of emission into the guided modes $I_{\mathrm{guided}}$, and show the results in Fig. \ref{fig4} (solid curves). 
For comparison, we also show the results for a single atom (dashed curves).
The comparison between the solid and dashed curves shows that the energy emitted into the guided modes, determined by the area under the intensity curve, is enhanced by the collective effect. 
Typical features of superradiance, such as
the increase of the emission rate, the occurrence of a local peak, and the enhancement of the peak intensity, are observed. However, they are rather weak in the case of Fig. \ref{fig4}. The reason is that the number of atoms $N$ is not large and the transfer rate $\gamma_{\mathrm{guided}}$ is small compared to the radiative decay rate  $\gamma_{\mathrm{rad}}$. We expect that the use of larger values for $N$
would lead to more dramatic effects. However, it is difficult to find the exact numerical solution to Eq. (\ref{13a}) for large $N$ because the number of variables is large.

In order to get insight into the case of large $N$, we make an approximation for the last term in Eq. (\ref{23}). 
For the initial product state $|\Psi\rangle$, we have $\langle\sigma_{j}^{\dagger}\sigma_{j'}\rangle=P(N-P)/N^2$ for every pair $j\not=j'$. We assume that this relation is valid for the whole emission process 
\cite{Eberly,Mandel book}.  
Such an assumption is reasonable under the condition $N\gg N-P_0\gg 1$ \cite{Gross}. 
With this assumption, Eq. (\ref{23}) yields
\begin{equation}\label{27}
\frac{d P}{dt}=-P[\gamma+(1-N^{-1})\gamma_{\mathrm{guided}}(N-P)].
\end{equation}
The solution to the above equation, subject to the initial condition $P(0)=P_0$, is 
$P=N{(\kappa+1)}/{[\kappa+e^{\Gamma (t+t_a)}]}$,
where $\kappa=(N-1)\gamma_{\mathrm{guided}}/\gamma$
and $t_a=\tau\ln[(\kappa+1)(N/P_0)-\kappa]$, with $\tau=\Gamma^{-1}$.
The intensity of emission into the guided modes is 
\begin{eqnarray}\label{44}
I_{\mathrm{guided}}&=&\hbar\omega_0 N \frac{\kappa+1}{\kappa+e^{\Gamma (t+t_a)}}
\nonumber\\&&\mbox{}
\times\left[\gamma(\kappa+1)\frac{e^{\Gamma (t+t_a)}}{\kappa+e^{\Gamma (t+t_a)}}-\gamma_{\mathrm{rad}}\right].
\end{eqnarray}
If $t_a<t_p$, where
$t_p=\tau\ln\{(1-N^{-1})[2+(N-2)(\gamma_{\mathrm{guided}}/\gamma)]\}$,
then the intensity $I_{\mathrm{guided}}(t)$ has a local peak with the height
$I_{\mathrm{guided}}^{\mathrm{max}}=\hbar\omega_0\gamma_{\mathrm{guided}} N^3/[4(N-1)]$
at the time $t=t^{\mathrm{max}}\equiv t_p-t_a$. Otherwise, the function $I_{\mathrm{guided}}(t)$ monotonically decreases from its initial value 
$I_{\mathrm{guided}}(0)=\hbar\omega_0P_0\gamma_{\mathrm{guided}}[1+(N-1)(1-P_0/N)]$.
Very similar features
have been obtained in the case of atoms in free space 
\cite{Eberly,Mandel book}.

\begin{figure}[tbh]
\begin{center}
  \includegraphics{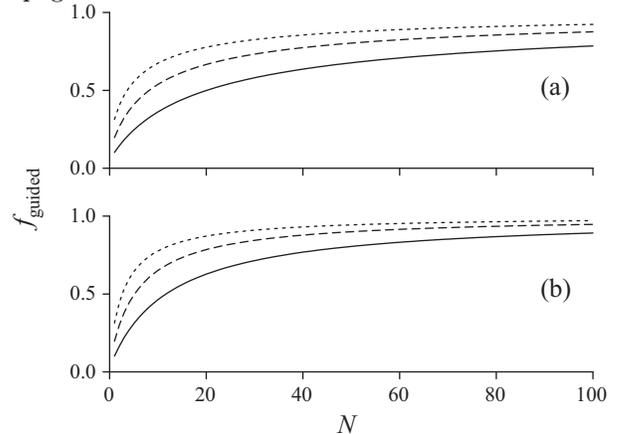}
 \end{center}
\caption{
Fraction $f_{\mathrm{guided}}$ of energy emitted from atoms initially prepared in a product state into the guided modes as a function of the atom number $N$. The angle $\theta$ for
the initial product state is (a) $\theta=0$ (full excitation) and (b) $\theta=\pi/2$ 
(one-half excitation). The atom--surface distance is $r-a=200$ nm (solid line), 100 nm (dashed line), and 0 (dotted line). The parameters used are as in Fig. \ref{fig2}.
}
\label{fig5}
\end{figure}

It follows from Eq. (\ref{44}) that the energy emitted into the guided modes is   
$U_{\mathrm{guided}}=\hbar\omega_0 P_0\{1
-\frac{N\gamma_{\mathrm{rad}}}{P_0\gamma}\frac{1}{\kappa}
\ln\frac{\kappa+1}{1+(1-P_0/N)\kappa}\}$.
Meanwhile, the total emitted energy is $U=\hbar\omega_0 P_0$. Hence the fraction of energy emitted into the guided modes is
\begin{equation}\label{50a}
f_{\mathrm{guided}}=1
-\frac{N}{P_0}\frac{\gamma_{\mathrm{rad}}}{\gamma}\frac{1}{\kappa}
\ln\frac{\kappa+1}{1+(1-P_0/N)\kappa}.
\end{equation}
Under the condition $P_0\cong N$, we have
\begin{equation}\label{52}
f_{\mathrm{guided}}=1-\frac{\gamma_{\mathrm{rad}}}{\gamma}\frac{\ln(\kappa+1)}{\kappa}.
\end{equation}
Equation (\ref{52}) together with the expression $\kappa=(N-1)\gamma_{\mathrm{guided}}/\gamma$
indicate that $f_{\mathrm{guided}}$ increases with increasing $N$ and that
$f_{\mathrm{guided}}\to 1$ in the limit $N\to\infty$ (see Fig. \ref{fig5}).

Since the propagation effects are neglected in our model, 
the above results are valid only if a photon can traverse the sample in a time shorter than the characteristic time scale of the collective decay. Therefore, the length $L$ of the atomic string in our model is limited  by the condition $L\ll L_0$, where $L_0 = c/\Gamma$ is the cooperativity length.
When we take $N=100$ and $\gamma_0=5.3$ MHz, and use the parameters $\gamma_{\mathrm{guided}}=0.26\gamma_0$
and $\gamma_{\mathrm{rad}}=1.06\gamma_0$, obtained in Fig. \ref{fig2} for the atom--surface distance $r-a=100$ nm, 
we find $L_0=33$ cm. 
For $L\gtrsim L_0$, the collective effects can still survive but the propagation effects must be included.

In conclusion, we have shown the possibility of directional \textit{guided} superradiance from a string of atoms that are separated by one or several wavelengths in a line parallel to the axis of a nanofiber. The rate of emission is enhanced by the cooperativity of the atoms via the guided modes. The efficiency of channeling of emission into the guided modes increases with increasing atom number and approaches unity in the limit of large numbers of atoms. In particular, for a string of 100 atoms prepared in the symmetric one-excitation state at the 100-nm distance from the surface of a 200-nm-radius fiber, the channeling efficiency can be as high as 96\%.

This work was carried out under the 21st Century COE program on ``Coherent Optical Science.''


\end{document}